# On-chip omnidirectional electromagnetic-thermal cloak

Yichao Liu[1], Hanchuan Chen[1], Gang Zhao[1], and Fei Sun[1,*]

*1 Key Lab of Advanced Transducers and Intelligent Control System, Ministry of Education and Shanxi Province, College of Electronic Information and Optical Engineering, Taiyuan University of Technology, Taiyuan, 030024 China*

\* sunfei@tyut.edu.cn

**Abstract**

Simultaneously guiding electromagnetic waves and heat flow at any incidence angle to smoothly bypass some electromagnetic/thermal sensitive elements is a key factor to ensure efficient communication and thermal protection for an on-chip system. In this study, an omnidirectional on-chip electromagnetic-thermal cloak is proposed. Firstly, a holey metallic plate with periodic array of subwavelength apertures is designed by optical surface transformation to realize an omnidirectional electromagnetic cloaking module for on-chip electromagnetic signal. Secondly, a two-layer ring-shaped engineered thermal structure is designed by solving Laplace equation to realize an omnidirectional thermal cloaking module for in-chip heat flow. Finally, these two cloaking modules are combined elaborately to achieve cloaking effect for both the electromagnetic waves and thermal fields simultaneously from any detecting direction, thus protecting the build-in electromagnetic/thermal sensitive elements without disturbing the external electromagnetic/thermal signal. The proposed electromagnetic-thermal cloak may have potential advantage in dealing with omnidirectional electromagnetic compatibility/shielding and multi-directional thermal management/dissipation of an on-chip system.

**Introduction**

Electromagnetic (EM) compatibility and thermal management are two key issues that needed to be addressed in our increasingly intelligent life, e.g., advanced driver assistance system, where strictly closed internal components need good thermal dissipation and electromagnetic interference (EMI) shielding conditions [1,2]. In fact, for almost all on-chip circuits and intelligent systems, the above problems should be confronted and resolved. Therefore, researchers have been working to find new composite materials with good thermal conductivity and EMI shielding properties [3-5]. With the development of metamaterials [6-9], many novel artificial structures have gained more advantage for controlling EM waves and thermal fields against traditional composite materials as they own exotic parameters that do not exist in natural materials, e.g., negative indices/conductivities [10-13], near-zero indices/conductivities[14-16], or extreme anisotropic properties [17,18], which has been widely used in the design of novel EM/thermal devices, often separately. Among them, optical or thermal null media [19-21] have extreme anisotropic permittivity, permeability, or thermal conductivity, allowing them to guide EM waves and heat flow propagating along a pre-designed path

(i.e., the principal axes of the null medium) in a highly directional manner. The functionality of the null medium can be easily understood by optical/thermal surface transformation [22,23], where a single surface (or two closely adjacent surfaces) in the reference space is separated apart by an extreme stretching transformation, forming the input and output surfaces of the null medium in the physical space. As the input and output surfaces of the null medium in the physical space correspond to a single surface in the reference space, EM waves and thermal fields at the output surface remain the same amplitude and phase as the input surface. In previous studies, thermal null medium and optical null medium (ONM) are studied separately, and therefore the thermal fields and EM waves could only be controlled individually. In a recent study [24], thermal/electromagnetic null medium that work for both the EM waves and thermal fields are proposed and successfully used to design a bi-physical-fields unidirectional cloak. However, the unidirectional EM-thermal cloak only works under restricted incident angle and beam width. In this study, an on-chip omnidirectional EM-thermal cloak that can work at arbitrary incident angle and under unrestricted beam width is proposed, which will be an effective way to solve the omnidirectional EMI shielding and multi-directional thermal dissipation problems.

The problems encountered when working on a highly integrated on-chip structure without the presence of the omnidirectional on-chip EM-thermal cloak is shown in Fig. 1(a), where the black sheet represents the chip, and the central four structures represent four on-chip EM-thermal sensitive elements, e.g., central processing unit, thermal/humidity sensors, and resonator with protective covers. For highly integrated on-chip systems, some EM-thermal sensitive components (e.g., the central processor) have to be positioned very close to other on-chip elements (e.g., the loop antenna in Fig. 1, or some other heat-generating components). In this situation, from an EM compatibility perspective, the four on-chip EM-thermal sensitive elements may be influenced by the EM signals from nearby antennas, and the presence of these sensitive elements will disrupt the original radiation pattern of the antenna (see the destroyed yellow radiation pattern in Fig. 1(a)), thereby interrupting or even blocking EM communication of on-chip antennas. From an in-chip thermal management perspective, the four EM-thermal sensitive elements may also be influenced by the heat flow (indicated by red arrowed lines) generated by surrounding components (e.g., the antennas will appear as a heat-generating component externally, after it receives EM radiation due to a portion of the energy is converted into heat energy), which may enter into the four EM-thermal sensitive elements, heating them up or even burning them (see the red heat flows in Fig. 1(a)). Moreover, the presence of four EM-thermal sensitive elements can also impact the distribution of heat flow and thermal management within the on-chip system. For example, the original directional cold flow generated by the on-chip semiconductor cooler for directional cooling will experience "heat scattering" due to the presence of the four EM-thermal sensitive elements, which will lead to a poor cooling effect for the target area (i.e., antenna array in Fig. 1(a)). Therefore, to simultaneously achieve EM compatibility and effective thermal management of the highly integrated on-chip systems, it is necessary to cloak the on-chip sensitive elements for EM waves and heat flows simultaneously.

In this study, both EM and thermal cloaking modules are carefully designed and ingeniously combined to perform as an omnidirectional on-chip EM-thermal cloak, which are placed on the chip with the four EM-thermal sensitive elements inside the concealed regions. The basic schematic diagram for the omnidirectional on-chip EM-thermal cloak is shown in Fig. 1(b), where the orange hollow cylinder with slits is the EM cloaking module, and the blue/orange thin layers in the chip represent the thermal cloaking module. EM waves generated or received by surrounding EM antenna can tunnel through the slits in the EM cloaking module and totally transmitted without touching the EM-thermal sensitive elements in four petal-shaped concealed regions. Moreover, heat flow will be guided by the thermal cloaking module around the concealed regions without heating up the EM-thermal sensitive elements. Under the protection of the EM-thermal cloak, the four EM-thermal sensitive elements are not affected by external EM radiations and heat flows, maintain a comfortable temperature, do not interfere with each other, and do not interfere with external EM-thermal signals. Therefore, the proposed on-chip EM-thermal cloak can effectively address both omnidirectional EMI shielding and multi-directional thermal dissipation challenges in highly integrated on-chip systems simultaneously.

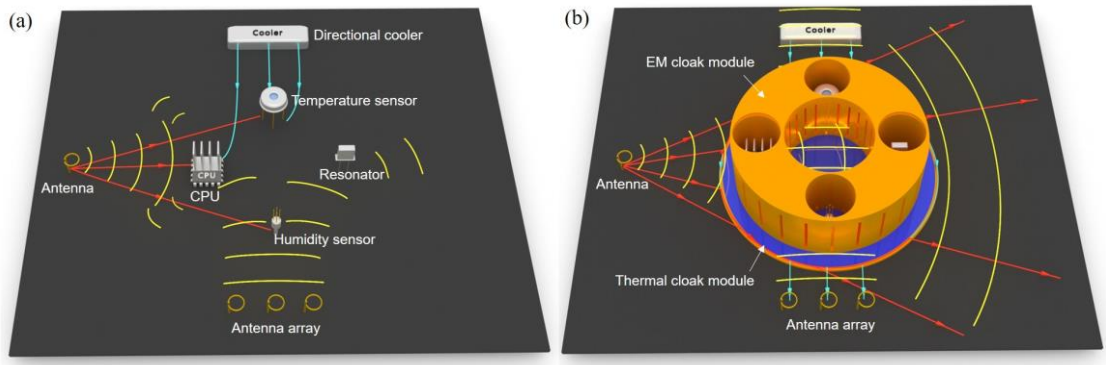

Fig. 1. Schematic diagram of an on-chip system involving omnidirectional EMI shielding and multi-directional thermal dissipation problems. (a) Without EM-thermal cloak: poor EMI shielding (e.g., the yellow EM radiation will produce EM interference on CPU and other on-chip elements), interrupted EM communication (e.g., the yellow EM signals produced by antenna array will be interrupted/blocked), and poor thermal management (e.g, the blue directional cold flow generated by the semiconductor cooler, originally flowing towards the antenna array, will be obstructed; the red heat fluxes generated by antenna will affect the operation of the nearby CPU). (b) With the EM-thermal cloak: good EMI shielding, high efficiency EM communication, and good thermal management.

**Design and implementation of the on-chip omnidirectional EM-thermal cloak**

For on-chip applications, we illustrate the design method of the omnidirectional EM-thermal cloak in the two-dimensional (2D) space. We firstly design two cloaking modules which are effective to EM wave and temperature field, respectively, by optical surface transformation and solving Laplace's equation. Secondly, we combine these two cloaking modules elaborately to form an on-chip omnidirectional EM-thermal cloak. Thirdly, three-dimensional (3D) EM metamaterials (i.e., holey metallic plate with periodic array of subwavelength apertures) and engineered thermal materials (i.e., thermal insulation layer and efficient heat conduction layer with radii satisfying a specific relationship) are designed, respectively, to realize the above theoretically-designed two cloaking modules. Thereafter, 3D numerical simulations are conducted to verify the performance of the designed on-chip EM-thermal cloak, which can guide the on-chip EM waves and in-chip heat flows around the common concealed region simultaneously.

We first show how to theoretically design EM cloaking module and thermal cloaking module, which can be divided into three steps (see Fig. 2). The first step is to design an EM concentrator using optical surface transformation [22,23]. The designed EM concentrator consists of a donut shape ONM shell with its inner radius $R_1$, outer radius $R_2$ and principal axes along the blue lines shown in Fig. 2(a), and a central core region with radius of $R_1$, which is filled with anisotropic magnetic material for TM-polarized EM waves, i.e., $\mu_r = diag(1,1,R_2^2/R_1^2)$ [25]. The second step is to create some concealed regions inside the ONM shell, which is shown in Fig. 2(b). By squeezing the principal axes towards the central line in each quadrant with the two ends fixed, four concealed regions colored black are created in Fig. 2(b). Due to the transformation invariance property of the ONM, the deformed ONM shell still has uniform parameters, only with the directions of the principal axes changed accordingly [26]. The deformed ONM can still guide rays (green curves in Fig. 2) around the concealed region without introduction of any reflections. The permittivity and permeability of both the original and deformed ONM shell can be written as $\varepsilon_r = \mu_r = diag(\infty, 0, 0)$, i.e., infinitely large along the principal axes and zero at directions perpendicular to the principal axes. The third step is to design thermal cloaking module by engineered thermal materials, which consist of a thermal insulation layer (colored blue) and efficient heat conduction layer (colored yellow) in Fig. 2(c). Thermal insulation layer is used to prevent external heat flow into the concealed region, and the efficient heat conduction layer is used to efficiently transport heat around the concealed region and the ONM shell, so that it can be quickly dissipated or recovered by a following cooler. To avoid introduction of interference to the external thermal flow, the thermal conductivity of the efficient heat conduction layer ($\kappa_c$) and background chip ($\kappa_b$) are restricted by [27]

$$\kappa_c = \kappa_b(R_4^2 + R_3^2)/(R_4^2 - R_3^2), \qquad (1)$$

where $R_3$ and $R_4$ are the inner and outer radius of the efficient heat conduction layer, respectively. The design diagram of above three steps is shown in Figs. 2(a)-2(c), respectively, and the corresponding 2D numerical simulation results for each step are given in Figs. 2(d)-2(f). Figs. 2(d) and 2(e) show the *z*-component of the normalized magnetic fields when TM-polarized planar EM detecting waves incident on the EM concentrator designed in the first step and on the EM cloaking module designed in the second step, respectively. Fig. 2(f) shows the distribution of the normalized temperature field when heat flow is incident onto the EM-thermal cloaking from the horizontal direction when the thermal cloaking module is added outside the EM cloaking module. The corresponding 2D simulation results with the ideal parameters in Figs. 2(d)-(f) verify that ideal EM cloaking module and thermal cloaking module can guide EM waves and heat flow smoothly around obstacles without any deflection, respectively.

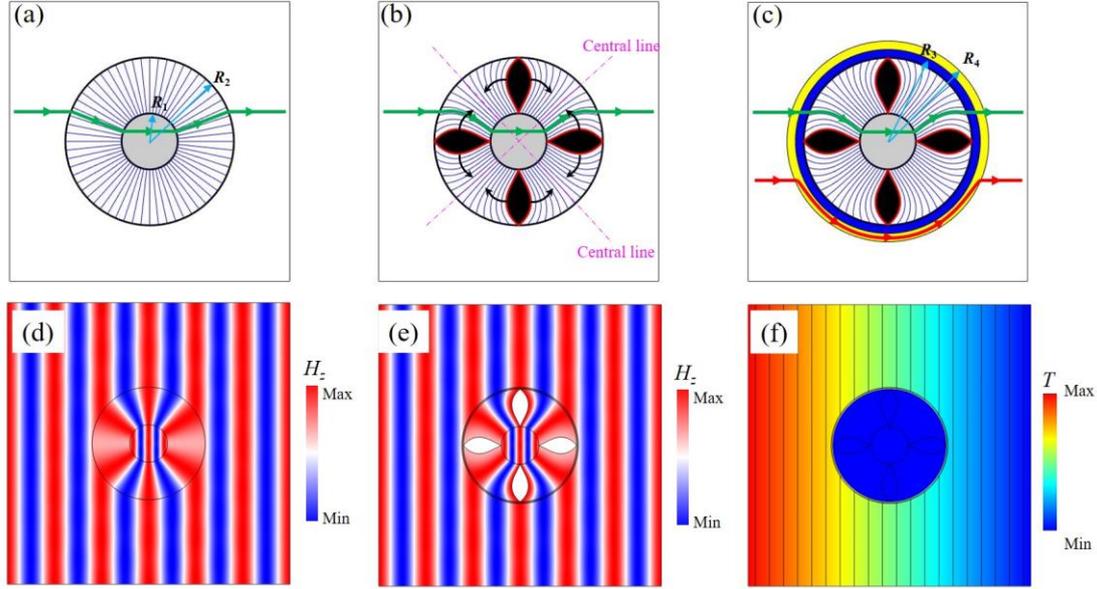

Fig. 2. (a)-(c) Basic design diagram of the EM-thermal cloak, where the green line and red line indicate light ray and heat flow, respectively. (a) Step 1: design a concentrator based on ONM; (b) Step 2: create four concealed regions inside the ONM to realize the EM cloaking module; (c) Step 3: design thermal cloaking module by engineered thermal materials, which consists of thermal insulation layer (colored blue) and efficient heat conduction layer (colored yellow) outside the ONM. (d)-(f) 2D simulated results. (d) Magnetic field distributions for an EM concentrator under illuminance of TM polarized plane wave. (e) Magnetic field distributions for the EM cloaking module under illuminance of TM polarized plane wave. (f) Temperature distributions for the EM-thermal cloak after the thermal cloaking module is added outside the EM cloaking module in (c). In these 2D simulations, the geometrical size of the EM-thermal cloaking is designed as $R_1$ =20mm, $R_2$ = 60mm, $R_3$ = 61mm, $R_4$ = 61.9mm. The material parameters are chosen as $\mu_r = 9, \varepsilon_r = 1$ for the magnetic core region, $\varepsilon_r = \mu_r = diag(\infty, 0, 0)$ in the principal coordinate system for the ONM shell, $\kappa_b = 6$ W/(m·K) for the background, $\kappa_c = 400$ W/(m·K) for the efficient heat conduction layer and $\kappa = 0$ W/(m·K) for the thermal insulation layer. In EM simulations (d)-(e), the frequency of the incident EM wave is $f_0$ = 6GHz, and surrounding boundaries are set as perfect matched layers. In thermal simulation (f), the left boundary is set at a fixed temperature 100°, the right boundary is set at a fixed temperature 0°.

Next, we will demonstrate how to implement the above EM-thermal cloak by 3D EM metamaterials and engineered thermal materials in a real on-chip structure, which is shown in Fig. 3(a). Firstly, we show the realization method of the upper EM cloaking module of the EM-thermal cloak. For the EM cloaking module, which is located above the on-chip structure for guiding EM waves above the chip around the concealed region, it consists of the ONM shell and magnetic core. There are several methods for the realization of ONM, such as staggered materials with positive and negative parameters based on the idea of complementary medium [28], zero index medium with doped metal [26], holey metallic plate with periodic array of subwavelength apertures [19], and Fabry–Pérot resonance-based metal channel structures [29]. For on-chip applications, a convenient but efficient method is using waveguide EM metamaterials, i.e., holey metallic plates. The 3D ONM shell is shown in Fig. 3(b), which is a hollow copper cylinder with air slit array. The petal-shaped regions are the concealed chambers. The x-y plane cross section of the 3D ONM shell is the same as the 2D case in Fig. 2(b), i.e., $R_2 = 3R_1 = 1.2\lambda_0$, where $\lambda_0 = 50$mm is the working wavelength. The height of the copper cylinder is set as $h = 0.7\lambda_0$. To achieve an effective ONM medium, the height of the slits inside the hollow cylinder is restricted to be half the working wavelength, i.e., $h_s = \lambda_0/2$, where the $TE_{10}$ mode is exactly at the cut-off frequency, and with vanished wave vector along the radial direction, thus resulting in the zero effective permittivity along tangential direction. The zero permeability along z-direction is obtained by restricting the slit width, which has a deep subwavelength scale and is selected as $w_o = \lambda_0/200$ on the outer surface and $w_i = \lambda_0/600$ on the inner surface in the present study. The periodicity of the slits along the tangential direction is set as $p_o = \lambda_0/8$ on the outer surface and $p_i = \lambda_0/24$ on the inner surface. The infinitely large permeability along the radial direction is naturally met as the copper can block microwaves propagating along the tangential directions. Therefore, the copper cylinder with air slit array has the same parameter as the ideal ONM, i.e., $\mu_\parallel = \infty$, $\mu_\perp = 0$, and $\varepsilon_\perp = 0$, for TM polarized EM waves. The core cylinder is magnetic material with permeability $\mu_r = diag(1, 1, 9)$, which can be realized by split ring resonator (SRR) arrays in Fig. 3(d). In our design, SRR unit is an 18-micrometer-thick split-ring-shaped copper sheet on a flexible printed circuit board with a thickness of 24 micrometers and a dielectric constant of 4.6. The geometric dimensions of SRR are designed as follows: $a = 2.2$mm, $r = 2.6$mm, $g = 0.28$mm, $b = 0.2$mm, $p = 6.67$mm. The effective permittivity and permeability of the SRR unit can be obtained through the S-parameter retrieval method [30], which are shown in Fig. 3(e). At the working frequency of 6 GHz, the SRR unit exhibits an effective permittivity of 1 and an effective permeability of 9, precisely meeting the requirements of the central magnetic core.

Secondly, we show the realization method of the lower thermal cloaking module of the EM-thermal cloak by engineered thermal materials. The thermal cloaking module is composed of a central thermal insulating cylinder (i.e., expanded polystyrene (EPS) with thermal conductivity of $\kappa_i = 0.04$ W/(m · K)) slightly higher than the surface of the chip and a ring-shaped efficient heat conduction layer (here we use copper with thermal conductivity of $\kappa_c = 400$ W/(m · K)) with the same height as the surface of the chip. To construct the whole EM-thermal cloak, the EM cloaking module (the orange 3D ONM shell and gray magnetic core) is placed on the thermal cloaking module, and the whole structure is embedded in the background chip (see the cross-sectional view of the EM-thermal cloak in Fig. 3(c)). Next, the central EPS cylinder with a radius of $R_3 = 61$ mm (slightly larger than the outer boundary of the EM cloaking module $R_2 = 60$ mm) and a height of $h_i = 2$ mm, and a ring-shaped copper layer with outer radius $R_4 = 61.9$ mm (calculated by inversely using Eq. (1)) and a height of $h_c = 1$ mm (the same as the background chip with thermal conductivity of $\kappa_b = 6$ W/(m·K) are used as an example in the following numerical studies.

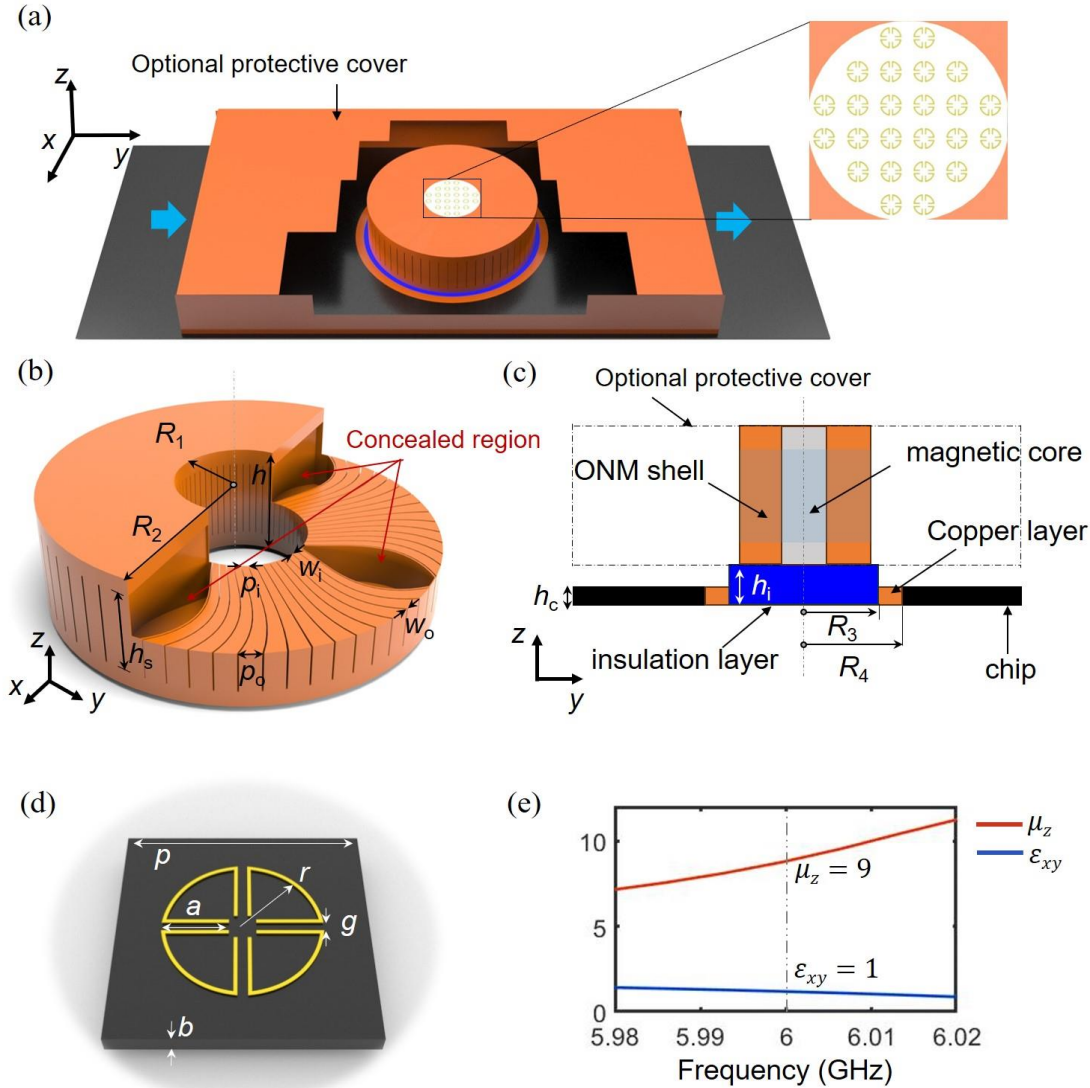

Fig. 3. (a) 3D structure of the EM-thermal cloak. To see clearly the structures inside, a hole is opened at the top and side metal layer, and the bottom metal layer is not shown. In practice, there is often a protective cover added above the on-chip structure, which can be modeled as an optional rectangular EM waveguide in simulations. (b) The detailed structure of the upper EM cloaking module of the EM-thermal cloak. (c) The cross-sectional view of the whole on-chip EM-thermal cloak in $y$-$z$ plane. (d) The schematic diagram of SRR unit cell for the realization of the magnetic core. (e) The relationship between frequency and the effective permittivity (blue line) or permeability (red line) of the SRR unit.

## Results and discussions

Now 3D numerical simulations are conducted to verify the cloaking effect of the on-chip omnidirectional EM-thermal cloak designed above for EM waves and thermal fields, respectively. All numerical simulations in this study are performed using COMSOL Multiphysics with the license number 9406999. The following simulations are all 3D cases in which the wave optics module and solid heat transfer module, with a steady-state solver, are chosen to simulate electromagnetic waves and thermal fields, respectively.

Firstly, we verify the cloaking effect of the designed EM-thermal cloak for EM waves numerically. Considering that on-chip systems are often enclosed by a protective cover in practice, we will simulate the cases when a protective cover is added above the on-chip structure (i.e., the optional cover in Fig. 3(a)) and when no protective cover is added (i.e., the on-chip structure is open to the air). If an optional protective cover is added above the on-chip system, it will perform as an equivalent metal waveguide for EM waves in Fig. 3(a), which is modeled by four metal layers in *z* and *x* directions. The 3D simulation results show that the designed EM cloaking module can create a good cloaking effect for EM waves when a protective cover is added above the on-chip system in Fig. 4(a) and 4(b), where the EM cloaking module in Fig. 4(b) has a rotation of 45 degree in the *x-y* plane compared with that in Fig. 4(a). Expected good EM cloaking effects are obtained for both incident angles. In contrast, if the designed EM cloaking module is removed, the four petal-shaped metal objects, which represents other metal elements components on the chip (e.g., instrumentation support structure, heat dissipation structure), will cause significant EM scattering and disrupt the original EM field distribution in the surrounding environment (see Figs. 4(c) and 4(d)).

If no optional protective cover is added above the on-chip system (i.e., the optional metal waveguide in Fig. 3(a) is removed), the designed EM cloaking module can still reduce the EM scattering from the four petal-shaped metal objects in the concealed regions. In this case, a magnetic line source (represents an on-chip integrated EM radiation antenna) in *z*-direction is placed very close (i.e., with a distance of $0.1R_2$) to the EM cloaking module, which can mimic the influence of the four petal-shaped metal objects on a nearby EM radiation antenna. Magnetic field distributions with cloak and without cloak are given in Fig. 4(e) and 4(f), respectively. If the EM cloaking module is introduced in Fig. 4(e), the four petal-shaped metal objects in the concealed regions have little influence on the EM radiation pattern and high transmittance of EM waves is achieved. However, large scattering occurs when the EM cloaking module is removed and results in a low transmittance of EM waves.

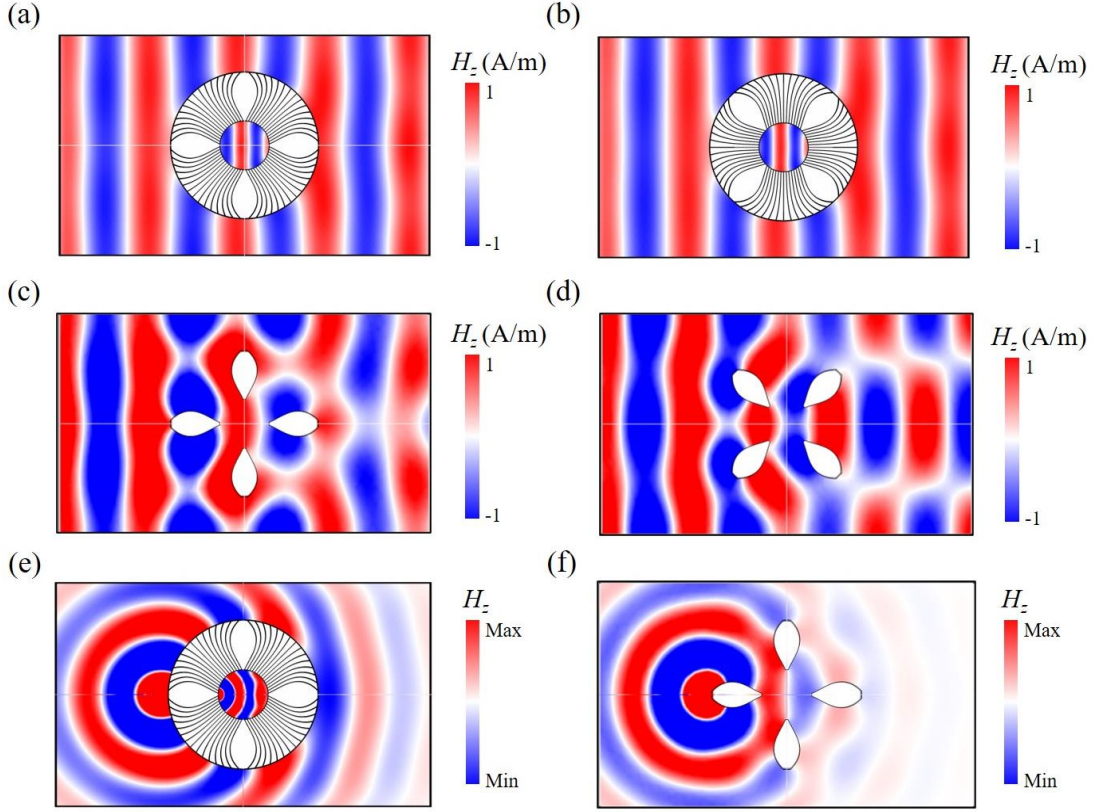

Fig. 4. 3D simulation results for the situation with (a-d) and without (e-f) the protective cover. (a) Magnetic field distribution of metal objects with cloak. (b) Magnetic field distribution of metal objects with a rotated cloak (rotated by 45 degrees in the *x-y* plane). (c) Magnetic field distribution of four bare metal objects. (d) Magnetic field distribution of rotated metal objects (rotated by 45 degrees in the *x-y* plane). (e) Magnetic field distribution of metal objects excited by a line magnetic source with cloak and without the protective cover. (f) Magnetic field distribution of metal objects excited by a line magnetic source without cloak and without the protective cover. In simulations, the optional protective cover is modeled as an EM metal waveguide with the input/output ports open, which has a cross section of $0.7\lambda_0$ (*z* direction) × $3.6\lambda_0$ (*x* direction). Fundamental mode is excited at the input port and propagate inside the EM waveguide with propagation constant $\beta = \sqrt{\left(\frac{2\pi}{\lambda_0}\right)^2 - \left(\frac{\pi}{h}\right)^2}$ . here, $h$ is the height of the optional protective cover, which is the same as the height of the ONM shell in Fig. 3(b).

    As shown in Fig. 4, to achieve on-chip EM cloaking effect, the EM cloaking module is essential. However, when only the EM cloaking module is present, it will also affect the in-chip heat flow distribution, which not only fails to achieve thermal cloaking for the four concealed regions where heat-sensitive components are present, but instead leads to an increase in the temperature of the four concealed regions and a significant perturbation of the temperature field around the on-chip system. To verify this, corresponding simulated results where an external heat flow directly incidents onto the EM cloaking module (without the thermal cloaking module) and a point heat source is

in front of the EM cloaking module (without the thermal cloaking module) are given in Fig. 5(b) and Fig. 5(d), respectively. The 3D simulated results in Fig. 5(b) and Fig. 5(d) show that only the EM cloaking module (without the thermal cloaking module) will cause significant disturbances to the background temperature field (i.e., "thermal scattering"), as well as significant warming of the concealed regions due to external heat flux. Therefore, the introduction of the thermal cloaking module is very necessary to achieve in-chip temperature field cloaking effect.

Next, we will numerically verify that the EM cloaking module together with the thermal cloaking module (i.e., the whole EM-thermal cloak in Fig. 3(a)) can eliminate the "thermal scattering" caused by the EM cloaking module on the in-chip heat flow and well achieves the thermal protection of the four concealed regions. If the whole EM-thermal cloak is introduced, the corresponding simulated results where an external heat flow directly incidents onto the cloak and a nearby heat source is in front of the cloak are given in Fig. 5(a) and Fig. 5(c), respectively. In this case, whether it is the heat flow impact or the influence of the surrounding high temperature heat source, after adding the thermal cloaking module, it not only plays a thermal buffer (cooling) effect on the whole cloak (i.e., the whole cloak as well as the sensitive elements inside the concealed regions can be thermally protected without being heat up), but at the same time does not have any influence on the background heat flow distribution (i.e., no "thermal scattering" appears).

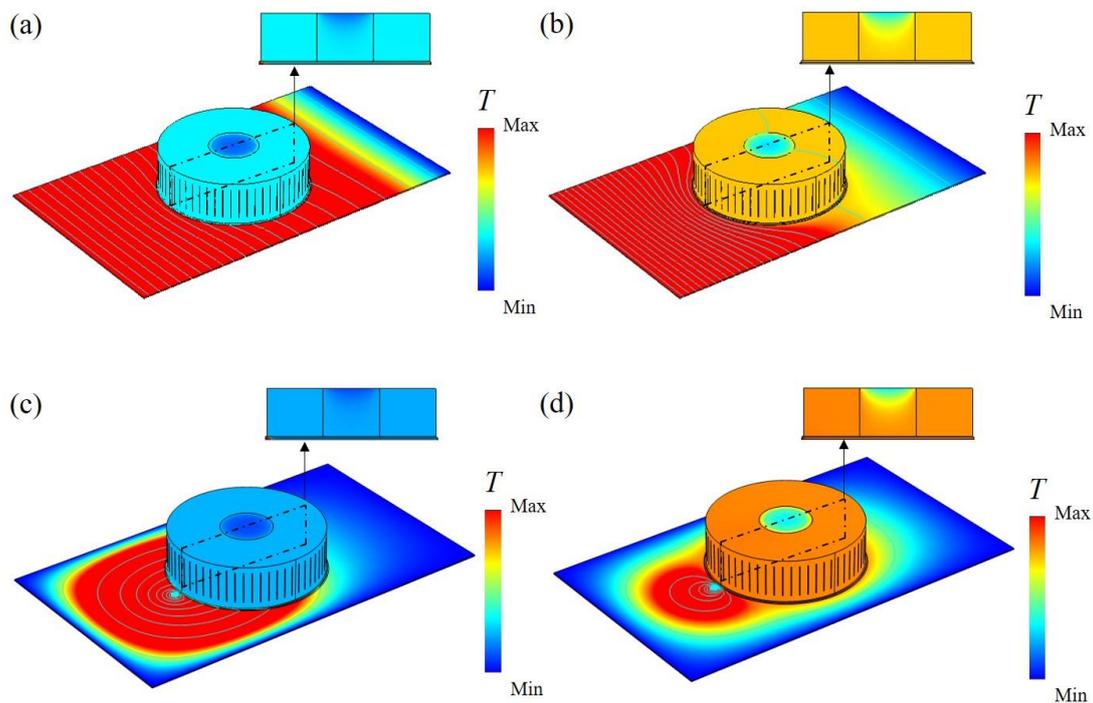

Fig. 5. 3D simulations depict two common thermal impacts on the thermal sensitive region of an on-chip system: (a) and (b) show directional thermal impacts from an external heat source, while (c) and (d) show thermal impacts from a nearby point source.

(a)-(b) Temperature fields distribution for (a) the EM-thermal cloak and (b) ONM shell directly placed on the chip, respectively. In these thermal simulations, the left boundary is set as a constant high temperature 70°C, the right boundary is set as a constant low temperature 20°C, the top and side boundaries of the ONM shell is set as thermal convection boundaries with heat transfer coefficient of 5W/(m$^2$·K) and outside room temperature of 20°C. (c)-(d) Temperature fields distribution for a nearby point thermal source (c) with EM-thermal cloak and (d) only with EM cloak. In this case, a thermal point source with fixed high temperature 70°C on the chip is placed very close (i.e., with a distance of $0.1R_2$) to the EM-thermal cloak, which can mimic the influence of the petal-shaped objects and EM cloak on a nearby thermal source. Constant room temperature of 20°C is used at the boundaries of the chip in (c) and (d). The temperature distributions inside the petal shaped object can be seen in the inset for each figure (cross-sectional view).

**Conclusion**

In conclusion, we design an EM cloak module for the on-chip EM waves and a thermal cloak module for the in-chip thermal flows, respectively, and then combine them elaborately to create an omnidirectional cloaking effect for both EM waves and thermal flows. The EM-thermal cloak has four concealed regions, and EM/thermal sensitive elements inside could get EMI shielding and be thermally protected simultaneously. Moreover, on-chip elements/components inside the proposed EM-thermal cloak will introduce no scattering/interference to external EM/thermal signals, which may solve the problems of electromagnetic compatibility and thermal management at the same time. Only natural materials (i.e., copper and EPS) are required for the proposed EM/thermal cloak, which make it more feasible for on-chip application.

**Data availability**

The main data and models supporting the findings of this study are available within the paper and Supplementary Information. Further information is available from the corresponding authors upon reasonable request.

**Acknowledgments**

This work is supported by the National Natural Science Foundation of China (Nos. 61971300, 12274317, 12374277, 61905208), and 2022 University Outstanding Youth Foundation of Taiyuan University of Technology.

**Competing interests**

The authors declare no competing financial interests.